# Simulated performance of a position sensitive radiation detecting system (COCAE)

K. Karafasoulis[1,2], K. Zachariadou[3,1*] S. Seferlis[1], I. Kaissas[1], C. Lambropoulos[4], D. Loukas[5], C. Poritiriadis[1]

[1]Greek Atomic Energy Commission, Patriarxou Grigoriou & Neapoleos P.O Box 60092, 15310, Athens, Greece

[2]Hellenic Army Academy, 16673 Vari, Greece,

[3]Technological Educational Institute of Piraeus, Thivon 250, 12244, Egaleo, Greece

[4]Technological Educational Institute of Chalkida, Psachna Evias, 34400 Greece

[5]Institute of Nuclear Physics, National Center for Scientific Research Demokritos, P.O Box 60228, Athens, Greece

*e-mail: zacharia@teipir.gr

**Abstract**

Extensive simulations of a portable radiation detecting system have been performed in order to explore important performance parameters. The instrument consists of a stack of ten detecting layers made of pixelated Cadmium Telluride (CdTe) crystals. Its aim is to localize and identify radiation sources, by exploiting the Compton imaging technique. In this paper we present performance parameters based on simulation studies. Specifically the ratio of incompletely absorbed photons, the detector's absolute efficiency as well as its energy and angular resolution are evaluated in a wide range of incident photon energies.

1.  **Introduction**

The COCAE instrument is a portable detecting system under development aimed to be used to accurately detect the position and the energy of radioactive sources up to ~2 MeV energy. Among its applications could be the security inspections at the borders and the detection of radioactive sources into scrap metals at recycling factories. It could also be used to improve the procedures at the nuclear waste management as well to provide fast and accurate information during the response after a nuclear emergency.

The COCAE instrument (Figure 1) consists of ten parallel planar layers made of pixelated Cadmium Telluride (CdTe) crystals occupying an area of 4cmx4cm, placed 2cm apart from each other. Each detector's layer consists of a two-dimensional array of pixels (100x100) of 400μm pitch, bump-bonded on a two-dimensional array of silicon readout CMOS circuits. Both pixels and readout arrays are on top of an $Al_2O_3$ supporting printed circuit board layer.

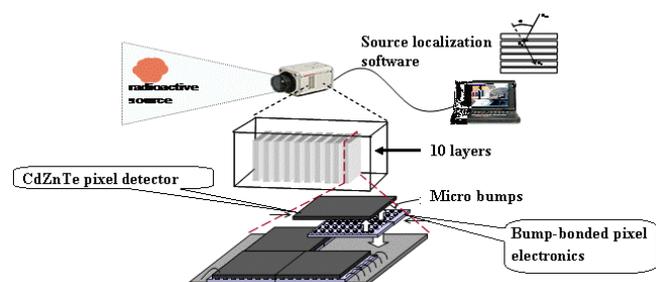

**Figure 1:** COCAE instrument conceptual design.

To detect the position of a radioactive source, the COCAE instrument exploits the Compton scattering imaging [1], a technique widely used in many fields such as nuclear medicine, astrophysics and recently counterterrorism.

When a photon interacts with the detector's sensitive materials both the energy deposition and the position of the interaction are recorded. The stored information is known as a "hit". If the photon interacts via the Compton scattering effect, a recoil electron and and a scattered photon are created. The energy and position of the recoil electron can be measured quickly, while the scattered photon, ideally, deposits all its energy in the detecting materials in a series of one or more interactions before it is finally absorbed via a photoelectric interaction. By measuring successive hits originating by Compton scattering of incident photons in the detecting CdTe layers, the incident direction of the primary gamma ray can be constrained to lie on a cone (Compton cone). The intersection of the cones corresponding to different incident photons determines the source location (see Figure .2). In principle, three cones are enough to reconstruct the image of a point source. In practice, due to measurement errors and incomplete photon absorption, many reconstructed cones are needed to derive the source location accurately.

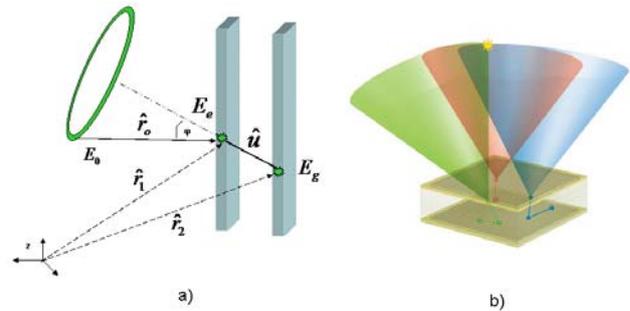

**Figure 2**: The principle of Compton reconstruction a) kinematics of Compton scattering b) Illustration of Compton imaging technique

The proposed instrument made of stacked CdTe semiconductor crystals, is expected to achieve an enhanced detection efficiency compared to Germanium (Ge) and Sodium Iodide (NaI) detectors. Furthermore, it is expected to have substantially better energy resolution compared to NaI detectors. Its challenge will be to achieve an energy resolution comparable to the one of high purity Ge detectors without requiring cryogenic cooling, thus being an accurate portable instrument.

Presented in this work are the results of extensive simulations studies of the COCAE instrument. The main objective of which is to explore various important performance parameters such as the response of the instrument under different radiation fields, the performance limiting factor of incompletely absorbed photons, the instrument's detecting efficiency as well as its energy and angular resolution.

## 2. Monte Carlo simulation description

The COCAE instrument is modelled by an open-source object-oriented software library (MEGAlib [2]) which provides an interface to the Geant4 [3] toolkit that simulates the passage of particles through matter. Within Geant4, the package G4LECS [4] is used to accurately model the Compton scattering by considering the influence of the Doppler broadening effect. The Monte Carlo simulation study encompasses the following steps:

A) The exact COCAE detector geometry is modelled. Compton interactions can occur not only in sensitive, but also in non-sensitive materials of the detector. Since this results to incomplete energy measurements (see section 3.1), special care has been taken to incorporate an accurate geometric and physical description of the detector's passive materials.

B) A point gamma ray source with isotropic emission is modeled, placed on the detector's axis of symmetry at 40cm from its first detecting layer. The source emits a large number of photons (~$2 \times 10^9$) in an energy varying from 60keV up to 2000keV.

C) The energy deposition of each photon with the detector ("hit") is recorded. The accurate simulation of the detector response is ensured by taking into account all relevant physical processes (Compton scattering, photoelectric effect, pair production, electron/positron transportation into matter, ionization). During the event reconstruction, dedicated algorithms are used in order to form events by using the spatial and energy information of each individual hit.

It has been already mentioned that COCAE instrument identifies the position of a radioactive source by reconstructing the direction of incident photons that interact via the process of Compton scattering. Multiple Compton scattering can occur in the detector's sensitive materials before the scattered photon is ideally fully absorbed in the detector's volume. Since the distance between the detecting layers is too short, it is impossible to have a time tag for each Compton interaction. This results to an ambiguity in the ordering of the Compton interactions sequence, which largely affects the correct determination of the Compton cone. Different techniques that exploit the kinematical and geometrical information of Compton scattering events as well as statistical criteria, have been extensively studied in order to select the best performing one ([5],[6]). The best algorithm reaches an efficiency of ~ 70% for incident photon energies above 600keV.

## 3. Detector characteristics

### 3.1 Energy spectrum

Due to the interactions of photons with the detector material, energy is transferred from the incident photon to the detector's electrons. The energy spectrum of the deposited energy in the detector's active material is a crucial parameter to understand the detector response under different radiation fields, thus extensive simulations have been performed. Shown in Figure 3 is the simulated energy spectrum, of different mono-energetic incident photons having energy that varies from 200 keV up to 1500 keV. The spectrum is obtained by summing the deposited energies in the sensitive materials of the detector. Its shape depends strongly on the mechanism via which the incident photon primarily interacts with the detector. If the primary photon interaction is a photoelectric effect, its energy is fully absorbed and it contributes to the full energy peak (photo-peak) of the energy spectrum. In contrast, a primary Compton interaction creates a scattered electron that carries only a fraction of the initial photon energy and a scattered photon that carries the remaining energy. If the latter is absorbed by a sensitive material of the detector, the event contributes to the photo-peak of the spectrum. Otherwise, the event contributes to the plateau at energies below the photo-peak (Compton plateau). Only Compton scattering events that are fully absorbed are useful for the source localization task. It is clear that the number of incompletely absorbed events (off-peak part of the energy spectrum) increases compared to the photo-peak events as the incident photon energy increases.

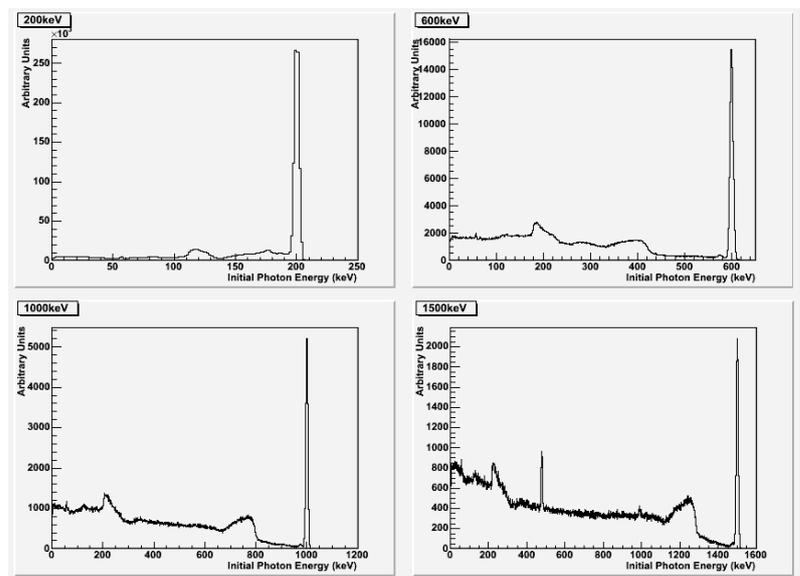

**Figure 3**: Simulated energy deposition spectra for different incident gamma ray energies.

The fraction of incompletely absorbed photons has been studied further by simulating a parallel photon beam placed on the detector's axis of symmetry. Figure 4 shows a) the energy spectrum of photons that do not interact with the sensitive or passive parts of the detector *("escaped energy")* and b) the energy spectrum of photons that deposit energy in the passive materials of the detector *("passive energy"), for 600keV incident photon energy.*

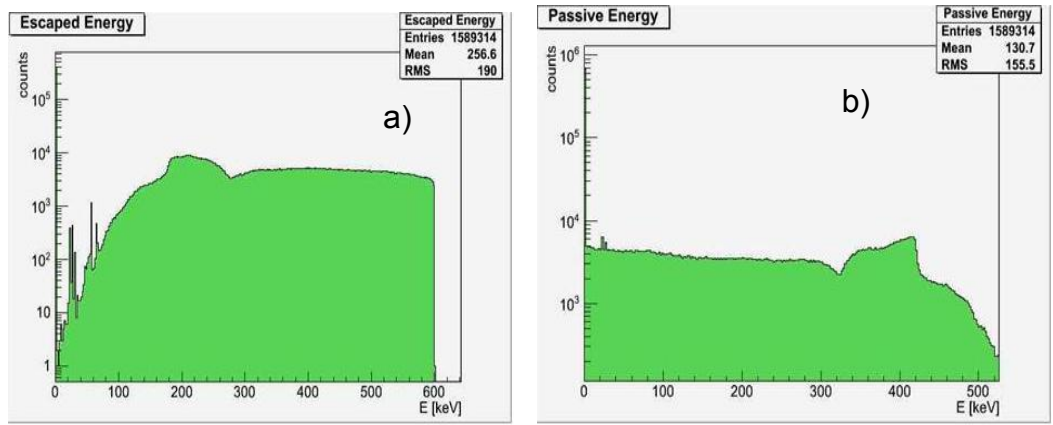

**Figure 4**: Energy spectrum for the a) "escaped" and b) "passive" energy

It can be noticed that the escaped energy spectrum is shifted towards higher energies whereas the opposite is true for the passive energy spectrum. This is expected since the probability that the photons escape from the detector is higher for higher energies.

The percentage of events having their first two interactions in the sensitive part of the detector is shown in Figure 5. In the same figure the percentage of events having their first interaction in the passive materials and the second one in sensitive materials of the detector is also shown. It is obvious that the latter event sample can not be used for the reconstruction of Compton events.

The *"passive energy"* and the *"escaped energy"* have been extensively studied for different inter-layer distances. It can be noticed (Figure 6) that the *"passive energy"* varies within 5% whereas the *"escaped energy"* varies up to 20% as the interlayer distances varies from 0.5cm to 2cm. Although the best case corresponds to 0.5cm, an interlayer distance of 2cm has been chosen for technical reasons.

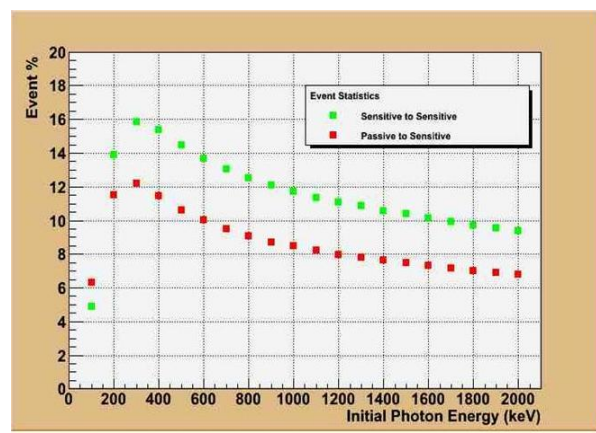

**Figure 5:** Event interactions in passive and sensitive materials of the detector

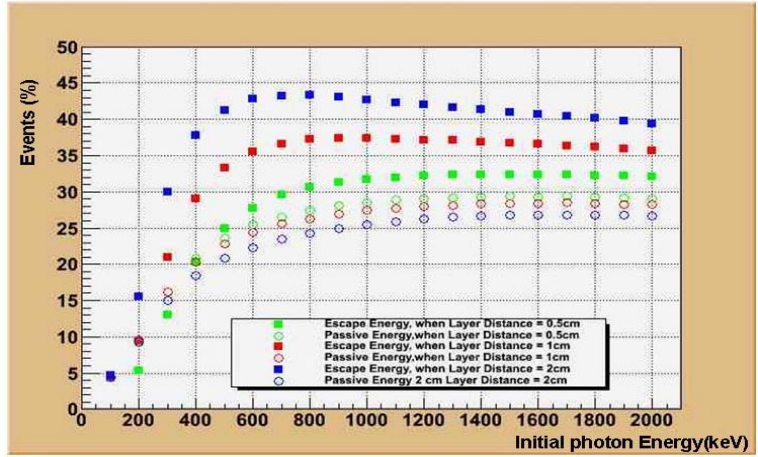

**Figure 6:** "passive" and "escaped" energy versus the initial photon energy

### 3.2 Detecting efficiency

The absolute detecting efficiency of the detector is defined as the fraction of recorded events in the photo-peak over the total number of generated events. It has been studied as a function of the incident photon energy, by modeling a point source placed at 71 cm from the first detecting layer.

Among the benefits of the COCAE instrument is its variable efficiency curve: It is possible to optimize the detection efficiency and the minimum detection limit imposed by the Compton plateau by activating only a part of its detecting layers. For example, if the identification of low energy gamma rays is of interest (like $^{241}$Am, $^{109}$Cd or the low energy part of the plutonium isotopes spectrum) the Compton plateau can be reduced by activating only one or two detecting layers of the instrument. Such a choice reduces the minimum detection limit and increases the sensitivity of the detector. Figure 7 shows the instrument's detecting efficiency for different number of activated detecting layers.

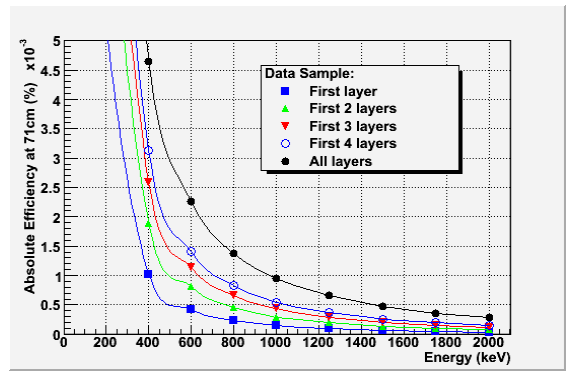

**Figure 7:** Detector's absolute efficiency for different number of activated detector's layers.

### 3.3 Energy resolution

The ability of the COCAE instrument to accurately localize radioactive sources depends strongly on the accuracy of the energy depositions measurements.

To model the realistic energy resolution of the COCAE instrument, the simulated energy depositions have been smeared according to a Gaussian distribution. Its FWHM has been determined using data from real measurements [7] taken by a mono-crystal CdTe Schottky diode, which is under development within the COCAE project, illuminated by various radioactive sources.

Presented in Figure 8 is the reconstructed energy resolution of the COCAE model for different incident photon energies. It is the convolution of the smeared energy depositions according to the real measurements, the Doppler broadening effect and the missing energy.

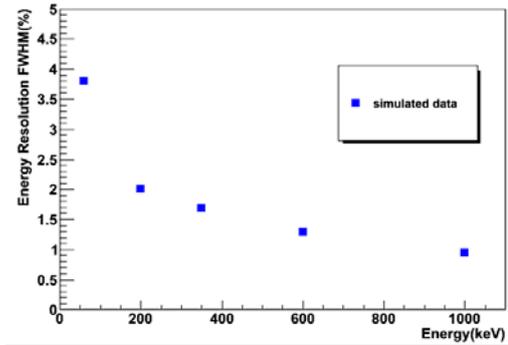

**Figure 8:** Reconstructed energy resolution as a function of the incident gamma energy

### 3.4 Angular resolution

To estimate the ability of the detector to evaluate the initial photon direction the Angular Resolution Measure (ARM) is used. The ARM is defined as the angle difference ($\Delta \phi = \phi^{kin} - \phi^{geo}$) between (a) the photon scatter angle $\phi^{kin}$ calculated by the measured energy depositions of the recoil electron and the scattered photon according to the Compton scattering kinematics and (b) the photon scatter angle $\phi^{geo}$ calculated by the positions ($\vec{r}_1$, $\vec{r}_2$) of the interactions and the direction $\hat{r}_0$ of the incident gamma ray (Figure 9)

The angular resolution of the detector is determined by the FWHM of the ARM distribution that has a well-shaped peak [6] (a convolution of a Gaussian and a Lorentz distribution) for low energy photons.

Figure 9 shows the angular resolution as a function of incident's photon energy for different detector energy resolutions. The lower limit (0% for a detector with perfect energy resolution) is set by the Doppler broadening effect. It can be notice that the angular resolution becomes better with increasing incident photon energy.

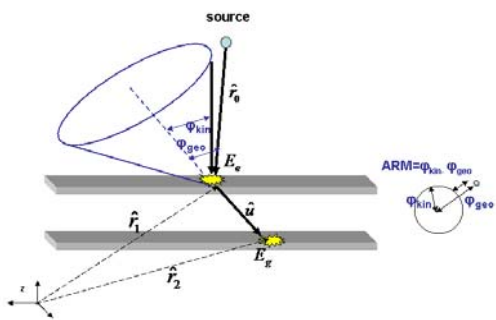

**Figure 9:** Sketch for the definition of the Angular Resolution Measure (ARM)

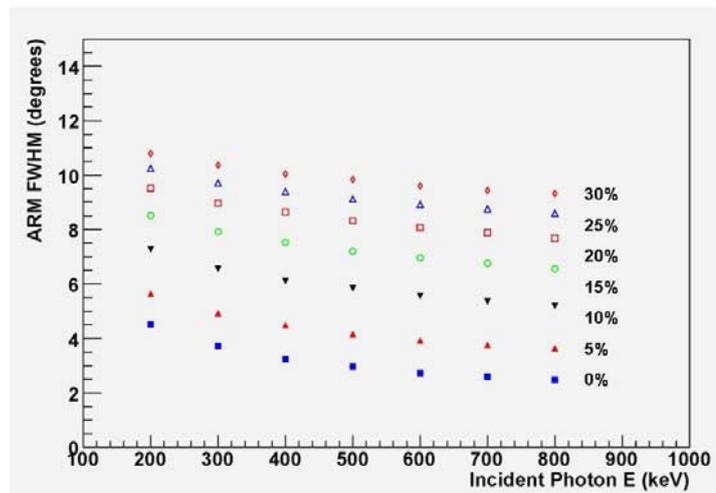

**Figure 10:** The detector angular resolution as a function of the incident photon energy, for different energy detector layer resolution

## 4. Discussion

In the present work, simulation studies of a portable multilayer position sensitive detector (COCAE) under development have been presented. Details of the simulation results of important detector performing parameters have been discussed. Specifically, the energy spectra of simulated gamma sources have been reconstructed and the limiting performance factor of the missing photon energy has been evaluated. In addition the detecting efficiency of the system has been estimated.

The reconstructed energy resolution is around 1% at energies above 600keV at room temperature. The angular resolution has been studied as a function of the incident gamma energy. For the case of 600keV incident photons is expected to be less than $4^0$ if the detector energy resolution is around 1%.

Towards the development of the COCAE instrument, apart the detailed simulation studies, two prototype integrated circuits have been developed, the pixelization of CdZnTe 75mm wafers has been successfully implemented and diode type single electrode detectors with high energy resolution have been developed. Furthermore, experiments with a precursor detector setup consisting of a stack of 3 CdTe hybrid pixel detectors are underway [7].